\documentclass[sigplan,twocolumn]{acmart}

\usepackage{tikz}
\usetikzlibrary{arrows.meta, positioning, fit, backgrounds}
\usepackage{subcaption}
\usepackage{bm}
\usepackage{enumitem}
\usepackage{mathtools}
\usepackage{pgfplots}
\pgfplotsset{compat=1.18}
\copyrightyear{2026}
\acmYear{2026}
\setcopyright{cc}
\setcctype[4.0]{by}
\acmConference[PaPoC '26]{Workshop on Principles and Practice of Consistency for Distributed Data}{April 27--30, 2026}{Edinburgh, Scotland Uk}
\acmBooktitle{Workshop on Principles and Practice of Consistency for Distributed Data (PaPoC '26), April 27--30, 2026, Edinburgh, Scotland Uk}
\acmDOI{10.1145/3806077.3806691}
\acmISBN{979-8-4007-2637-8/2026/04}

\begin{document}

\title{ERA: Epoch-Resolved Arbitration\texorpdfstring{\\}{} for Duelling Admins in Group Management CRDTs}

\author{Kegan Dougal}
\email{kegan@element.io}
\affiliation{
  \institution{Element Creations Ltd}
  \city{London}
  \country{United Kingdom}
}

\begin{abstract}
  Conflict-Free Replicated Data Types (CRDTs) are used in a range of fields
  for their coordination-free replication with strong eventual consistency.
  By prioritising availability over consistency under partition, peers accumulate events in different orders,
  and rely on an associative, commutative and idempotent merge function to present a materialised view of the CRDT.
  Under some circumstances, the state of the materialised view over time can appear to ``roll back'' previously applied events.
  When the materialised view is used to manage group permissions such as ones found in instant messaging
  applications, this can lead to surprising behaviour.
  Rollbacks can occur when there are multiple concurrent events, such as in the
  Duelling Admins problem where two equally permissioned admins concurrently revoke each other's permissions.
  Who wins?
  Different solutions and their trade-offs are examined.
  A Byzantine admin can exploit concurrency to influence the duel, whereby we argue that an external arbiter is required to order concurrent events.
  Our ERA proposal arbitrates asynchronously in batches via optional ``epoch events'', preserving availability.
  This introduces a bounded total order within epochs, and the resulting ``finality''
  improves on the level of consistency CRDTs can provide.
\end{abstract}


\maketitle

\newpage{}
\section{Introduction}
Conflict-Free Replicated Data Types (CRDTs)~\cite{shapiro2011conflict} are
in widespread use in a variety of fields including databases~\cite{riakDevelopingWith,akkaDistributedData},
consumer messaging~\cite{MatrixSpecificationV117} and collaborative apps~\cite{inkandswitchKeyhiveLocalfirst}.
Many products rely on a separate system for authorising access to the CRDT, e.g. OIDC / LDAP servers.
This can subvert the incentives~\cite{inkandswitchLocalfirstSoftware} for using a CRDT as it
reduces availability under partition should the authorisation system be unavailable.
Keyhive~\cite{inkandswitchKeyhiveLocalfirst} and Matrix~\cite{MatrixSpecificationV117} attempt to implement authorisation
based on Byzantine fault tolerant (BFT) CRDTs using hash DAGs~\cite{kleppmann2020byzantine} (also known as hash chronicles~\cite{jacob2024logical}, blocklaces~\cite{almeida2024blocklace}, and by other terms).
Whilst they implement decentralised access control differently~\cite{jacob2020,keyhiveGroupMembership},
both systems present to the application-level the ability to manage a group of users.
Users may be in different membership states such as \emph{invited}, \emph{joined}, or \emph{left}.
Users may also have different roles such as \emph{Administrator} which enables the ability to remove other users and change their roles.
In these systems, the group creator starts with the  \emph{Administrator} role and they can subsequently promote or demote
other users to this role. Demotion (or revocation) is problematic in these systems because it is
a non-monotonic operation~\cite{jacob2025knowledge}.
This means some previously authorised events may need to be \emph{rolled back} due to concurrent revocations.
This introduces the ``Duelling Admins'' problem: what should be the result of processing two concurrent revocations
where Admin A revokes Admin B whilst concurrently Admin B revokes Admin A?
This forms a revocation cycle which must be automatically resolved by the CRDT merge function.

Kleppmann discussed \emph{seniority ranking} as a solution to the Duelling Admins problem
at PaPoC 2025's keynote~\cite{kleppmannKeynoteByzantine}:
execute the more senior admin's operations first, e.g. A < B.
However, this prevents a less senior admin from ever revoking a more senior admin
because the less senior admin is unable to close the cycle.
Consequently, a revoked Admin A can retaliate against Admin B by \emph{backdating} a revocation
to make it appear as if they concurrently revoked each other.
This allows Admin A to \emph{roll back} their revocation, as seniority ranking does not support revoking a more senior admin.

In this article, we describe the Duelling Admins problem, explain how Keyhive and Matrix solve it, and explore how it can be solved in a fair way by the introduction of a mutually trusted peer called a \emph{finality arbiter}. This peer is responsible for
arbitrating the order of events via the transmission of epoch events.
We conclude with an outline of how CRDT designs may use the arbitration characteristics of finality arbiters in novel ways.

\section{Background}

CRDTs are updated by exchanging \emph{events} between all participating \emph{peers}.
These events are merged together by a \emph{merge function} to form a \emph{materialised view} of the CRDT.
The merge function must be associative, commutative, and idempotent in order to ensure all peers converge on the same
materialised view, regardless of the order in which they received the events from the network.
This presents a problem
because group membership operations are not commutative: in order for Alice to leave the group, she must already be joined to the group.
To handle this, both Keyhive and Matrix encode the causal past in the events they exchange. This encodes the happens-before
relation which is required in order to apply access control rules correctly.
The happens-before relation is often expressed via dotted version vectors~\cite{preguicca2010dotted}.
However, it is unsafe to use version vectors to represent causality in a Byzantine environment due to \emph{equivocation attacks}~\cite{jacob2021conflict}.
When two peers synchronise events in a CRDT, they typically employ event IDs as a compressed representation of the underlying event data.
Equivocation is the act of a Byzantine peer sending different event data to different peers, reusing the same event ID.
This results in convergence failures, as correct peers believe they are in-sync with each other.
Equivocation can be addressed with recursive hash linking:
an event ID is the hash of the event data with the event IDs of its causal predecessors.
Recursive hash linking forms a hash-linked Directed Acyclic Graph (hash DAG) of events, providing Byzantine eventual consistency~\cite{kleppmann2020byzantine}.

To calculate the materialised view of a DAG, the merge function is applied to the accumulated state at the DAG sources, i.e., the most recent events in causal order.
Because hash links point towards the past, the most recent events have no inbound links, i.e., are sources of the DAG.
Sources are called ``forward extremities'' in Matrix or ``heads'' in Automerge / Keyhive.
Only the sources need to be merged because every node in the DAG is an ancestor of at least one source, meaning all history is represented
by the CRDT state at those sources.
More formally, for a DAG $G = (V, E)$ and join function $\bigsqcup$, the materialised view $MV$ is:
\begin{align*}
\mathrm{Sources}(G)
&=\;
\{\, v \in V \mid \deg^{+}(v) = 0 \,\}
\\[0.5em]
\forall v \in V,\; \exists \ell \in \mathrm{Sources}(G)
&:\; v \preceq \ell
\\[0.5em]
\Longrightarrow\quad
\mathrm{MV}(G)
&=\;
\bigsqcup_{\ell \in \mathrm{Sources}(G)} \ell
\end{align*}

Fetching concurrent events fetches new DAG sources,
which may change the \emph{materialised view} of the CRDT,
which may \emph{roll back} previously valid group membership events.

\subsection{Finality and Causal Stability}

\begin{figure}[b]
\centering

\begin{subfigure}{0.45\linewidth}
\centering
\begin{tikzpicture}[
  node distance=1cm,
  msg/.style={
    draw,
    rounded corners,
    minimum size=5mm,
    inner sep=1.5pt
  },
  >=Stealth
]
  \node[msg] (a) {$a_1$};
  \node[msg] (b) [below right of=a] {$b_1$};
  \node[msg] (c) [below right of=b] {$a_2$};
  \node[msg,fill=gray!10] (d) [below left of=a] {$a_3$};

  \draw[<-] (a) -- (b);
  \draw[<-] (a) -- (d);
  \draw[<-] (b) -- (c);
\end{tikzpicture}
\caption{Detectable backdating: Peer $a$ backdated event $a_3$ which is detectable due to $a_3$ being concurrent with $a_2$.}
\end{subfigure}
\hfill
\begin{subfigure}{0.45\linewidth}
\centering
\begin{tikzpicture}[
  node distance=1cm,
  msg/.style={
    draw,
    rounded corners,
    minimum size=5mm,
    inner sep=1.5pt
  },
  >=Stealth
]
  \node[msg] (a) {$c_1$};
  \node[msg] (b) [below right of=a] {$d_1$};
  \node[msg] (c) [below right of=b] {$d_2$};
  \node[msg,fill=gray!10] (d) [below left of=a] {$c_2$};

  \draw[<-] (a) -- (b);
  \draw[<-] (a) -- (d);
  \draw[<-] (b) -- (c);
\end{tikzpicture}
\caption{Undetectable backdating: it is indistinguishable if peer $c$ backdated $c_2$ concurrent to $d_1$ or if $c_2$ is a late arriving event.}
\end{subfigure}

\caption{DAGs of causal histories demonstrating detectable vs. undetectable backdating. Each event points to its causal predecessor.
An event $x_i$ is the $i$th event of peer $x$.}
\label{fig:backdating}
\end{figure}

Finality and causal stability are mechanisms for guaranteeing that a given event cannot be rolled back.
\citeauthor{almeida2024approaches} defines \emph{causal stability}~\cite{almeida2024approaches} (for messages instead of events) as:

\begin{definition}[Causal Stability~\cite{almeida2024approaches}]
[An event] with timestamp $t$ is causally stable at peer $i$ when all [events] subsequently delivered at $i$ will have timestamp $t' \ge t$. 
\end{definition}

Timestamps are used in this definition to specify causality between events, which corresponds to event predecessors in the hash DAG.
In contrast, \emph{finality} encapsulates the desired property
that an event is ``final'' and cannot be \emph{rolled back} due to missed concurrent events.
In blockchains, this refers to the time when it becomes impossible to roll back the transactions in a block ~\cite{nakamoto2009bitcoin}.
Causal stability is a technique to determine finality in CRDTs by ensuring that there are no unknown concurrent events.
However, finality only requires that there are no concurrent events which would \emph{roll back} a given event $m$.

For concurrent events $m \parallel m'$, 
we say $m'$ \emph{rolls back} $m$ when $m$'s presence has no effect on 
the materialised view:
\begin{equation}
\mathsf{Rollback}(m', m)
\;\stackrel{\mathrm{def}}{\iff}\;
\mathrm{MV}(\{m, m'\}) \equiv \mathrm{MV}(\{m'\})
\label{eq:rollback}
\end{equation}

Finality is then the absence of any concurrent event that could cause a rollback:
\begin{equation}
\mathsf{Final}(m)
\;\stackrel{\mathrm{def}}{\iff}\;
\forall m' \colon\;
m' \parallel m
\;\Rightarrow\;
\neg\,\mathsf{Rollback}(m', m)
\label{eq:finality}
\end{equation}

Intuitively, Bob saying "hello" concurrently cannot affect whether Alice is joined, so her membership can be finalised without waiting for Bob's message.

\subsection{Backdating}

\textit{Backdating attacks} occur when a Byzantine peer sends an event to intentionally cause an event to be rolled back.
Backdating attacks fall into two categories: \textit{detectable} and \textit{undetectable}.
\textit{Detectable backdating} occurs when a peer sends an event concurrent with itself.
Let $p \in P$ be a peer, and let $E_p$ be the set of events sent by $p$:
\begin{equation}
\exists e_1, e_2 \in E_p \colon
\quad e_1 \parallel e_2
\end{equation}

To detect backdating, the protocol must forbid peers from sending concurrent events, meaning any concurrent event sent by the same peer is an attempt to backdate:
\begin{equation}
\forall n \in N,\; \forall e_1, e_2 \in E_n \colon
\quad \neg(e_1 \parallel e_2)
\end{equation}

\emph{Undetectable backdating} occurs when a peer sends an event concurrent to the latest received event, instead of causally after it.
This is undetectable because the peer is not sending an event concurrent to itself:
it is indistinguishable from a correct peer under poor network conditions who may genuinely not have received some events. Both forms are shown in Figure~\ref{fig:backdating}.

\section{Problem Statement}

Each peer represents a single \emph{user}. Consider a hash DAG of membership events. Each user has one of the following roles:
 \begin{enumerate}
  \item \label{role:read}
  \textbf{Reader:}
  The user can read events in the DAG. This is the lowest level role.
  \item \label{role:write}
  \textbf{Writer:}
  The user can write events into the DAG, in addition to the Reader role.
  \item \label{role:admin}
  \textbf{Admin:}
  The user can change the role of other users, in addition to the Writer role.
\end{enumerate}

The valid operations are as follows:
\begin{enumerate}
  \item \label{op:join}
  \textbf{$\mathsf{join}(a)$:}
  User $a$ joins the group. \emph{Any user} can do this, without explicit prior permission. The first user to join is an Admin. Subsequent users are a Reader.
  \item \label{op:write}
  \textbf{$\mathsf{write}(a)$:}
  User $a$ writes a message into the DAG. To do this, $a$ must be a Writer or Admin.
  \item \label{op:promote}
  \textbf{$\mathsf{promote}(a, b, \mathit{role})$:}
  User $a$ increases the role of user $b$ to $\mathit{role}$. To do this, $a$ must have permission by having the Admin role.
  \item \label{op:demote}
  \textbf{$\mathsf{demote}(a, b, \mathit{role})$:}
  User $a$ decreases the role of user $b$ to $\mathit{role}$. To do this, $a$ must have permission by having the Admin role, while $b$ may have any role.
  \emph{Self-demotions}, i.e., $a = b$, are valid.
\end{enumerate}

The following safety and liveness properties are required from the perspective of a single peer:
\begin{enumerate}[
  label=\textbf{P\arabic*:},
  ref=P\arabic*,
  leftmargin=*
]
  \item \label{prop:liveness}
  \textbf{Availability (Liveness).}
  Users are able to add events to their local DAG to be sent later even when they cannot communicate with \emph{any} other user.
  \item \label{prop:safety}
  \textbf{Authorisation Safety.}
  No user may successfully execute an operation requiring permissions they do not possess.
  \item \label{prop:impartial}
  \textbf{Concurrency Safety.}
  No user can influence the resolution of conflicts, even those they have manufactured.
\end{enumerate}

Liveness property ~\ref{prop:liveness} ensures that coordination-free replication is preserved.
Safety property ~\ref{prop:safety} ensures that the authorisation model, be it capabilities or access control lists, functions correctly.
Safety property ~\ref{prop:impartial} ensures that backdating cannot be used to gain an advantage in the system.

Consider the Duelling Admins problem, visualised in Figure~\ref{fig:dueladmins}.
What strategy should be used to comply with these safety and liveness properties?

\subsection{Diagram conventions}
\label{sec:conventions}
All diagrams represent a DAG, where nodes denote events and edges define hash DAG causal ordering.
The position of each node denotes its position in the \emph{execution order} when read from top to bottom,
indicating how concurrent events are sorted.
Events that are \emph{unauthorised} according to this order are indicated with $^{\bm{\times}}$.
Annotations to the right of each event show the effective role assignments visible \emph{after} the event is applied at
that point in the execution order.
Consequently, the \emph{materialised view} is the bottom-most annotation in each diagram.
\textbf{Bold text} marks events that are backdated.

\begin{figure}[b]
\centering
\begin{tikzpicture}[
  node distance=1cm,
  >=Stealth,
  note/.style={
    align=left,
    text width=3cm,
    font=\footnotesize,
  }
]
  \node (a) {$\mathsf{join}(a)$};
  \node (b) [below of=a] {$\mathsf{join}(b)$};
  \node (c) [below of=b] {$\mathsf{promote}(a, b, \mathsf{Admin})$};
  \node (d) [below left of=c, xshift=-7mm] {$\mathsf{demote}(a, b, \mathsf{Reader})$};
  \node (e) [below right of=c, yshift=-5mm] {$\mathsf{demote}(b, a, \mathsf{Reader})^{\bm{\times}}$};

  \draw[<-] (a) -- (b);
  \draw[<-] (b) -- (c);
  \draw[<-] (c) -- (d);
  \draw[<-] (c) -- (e);

  \coordinate (notes) at ([xshift=1.5cm] a.east);
    \node[note, anchor=west] at (notes |- a) {$a=\mathsf{Admin}$};
    \node[note, anchor=west] at (notes |- b) {$a=\mathsf{Admin}, b=\mathsf{Reader}$};
    \node[note, anchor=west] at (notes |- c) {$a=\mathsf{Admin}, b=\mathsf{Admin}$};
    \node[note, anchor=west] at (notes |- d) {$a=\mathsf{Admin}, b=\mathsf{Reader}$};
    \node[note, anchor=west] at (notes |- e) {$a=\mathsf{Admin}, b=\mathsf{Reader}$};

\end{tikzpicture}
\caption{The Duelling Admins problem. Which user remains an Admin depends on how $\mathsf{demote}$ events are sorted. Sorting by user centralises authority. Sorting by event ID causes repeated rollbacks.}
\label{fig:dueladmins}
\end{figure}
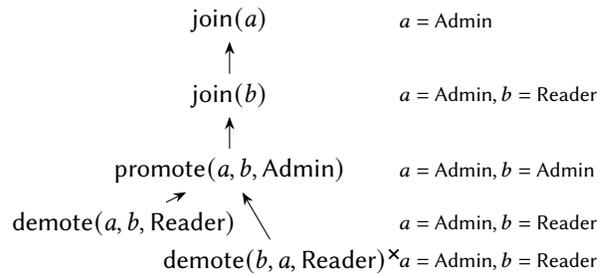

\begin{figure}[b]
\centering
\begin{tikzpicture}[
  node distance=1cm,
  >=Stealth,
  note/.style={
    align=left,
    text width=3cm,
    font=\footnotesize,
  }
]
  \node (a) {$\mathsf{join}(a)$};
  \node (b) [below of=a] {$\mathsf{join}(b)$};
  \node (c) [below left of=b, xshift=-7mm] {$\bm{\mathsf{demote}(a, a, \mathsf{Writer})}$};
  \node (d) [below right of=b, yshift=-5mm] {$\mathsf{promote}(a, b, \mathsf{Admin})^{\bm{\times}}$};

  \draw[<-] (a) -- (b);
  \draw[<-] (b) -- (c);
  \draw[<-] (b) -- (d);

  \coordinate (notes) at ([xshift=1.5cm] a.east);
    \node[note, anchor=west] at (notes |- a) {$a=\mathsf{Admin}$};
    \node[note, anchor=west] at (notes |- b) {$a=\mathsf{Admin}, b=\mathsf{Reader}$};
    \node[note, anchor=west] at (notes |- c) {$a=\mathsf{Writer}, b=\mathsf{Reader}$};
    \node[note, anchor=west] at (notes |- d) {$a=\mathsf{Writer}, b=\mathsf{Reader}$};

\end{tikzpicture}
\caption{Even if equal-permissioned roles are forbidden from demoting each other, backdating can subvert this. $\mathsf{demote}(a, a, \mathsf{Writer})$ executes first, meaning $a$ is no longer authorised to $\mathsf{promote}(a, b, \mathsf{Admin})$.}
\label{fig:equaldemote}
\end{figure}
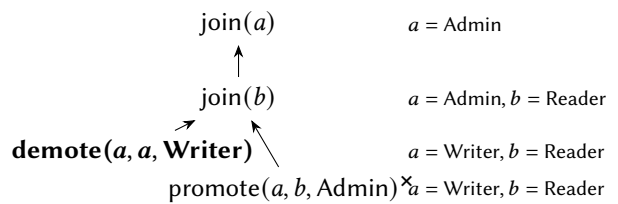

\subsection{Design Considerations}

CRDTs require the effects of concurrent events to be commutative.
However, authorisation is non-commutative:
To determine whether $\mathsf{demote}(b,a,\mathsf{Reader})$ is authorised, we must know whether $b$ is an $\mathsf{Admin}$ \emph{at the point of execution}, which depends on whether $\mathsf{demote}(a,b,\mathsf{Reader})$ has already been applied.
Each event's validity is contingent on the other not having executed first.
\emph{Arbitration}~\cite{almeida2024framework} is a global total order of events that all peers will eventually agree upon.
Revocation cycles can be broken via arbitration because arbitration orders events.
Without an arbitration, we must either accept both demotions, reject both, or pick one (which \emph{is} a form of arbitration).
Accepting or rejecting both demotions violates safety property ~\ref{prop:impartial} because a Byzantine user can influence the resolution by backdating a demotion.
Thus, we believe that authorisation conflicts should be resolved by \emph{arbitration}, acquired by peers deterministically sorting their local DAGs.

The protocol cannot forbid users with the same role from demoting each other because the user who first gained that role
can backdate, as shown in Figure~\ref{fig:equaldemote}.
This is detectable backdating, but excluding Byzantine users creates more problems.
Firstly, in this scenario, the \emph{intention} is to invalidate a promotion of another user, so excluding the promotion because it was sent by a Byzantine user is furthering the adversary's goal.
Secondly, in practice, users can accidentally backdate if they restore a database / mobile backup which can forget their most recently sent events.
Thus, based on our practical experience with operating Matrix, we do not see permanently excluding those users as proposed by Blocklace~\cite{almeida2024blocklace}, as practical.

\begin{figure}[h]
\centering
\begin{tikzpicture}[
  node distance=1cm,
  >=Stealth,
  note/.style={
    align=left,
    text width=3cm,
    font=\footnotesize,
  }
]
  \node (a) {$\mathsf{join}(a)$};
  \node (b) [below of=a] {$\mathsf{join}(b)$};
  \node (c) [below of=b] {$\mathsf{promote}(a, b, \mathsf{Writer})$};
  \node (d) [below left of=c, yshift=-5mm,xshift=-10mm] {$\mathsf{write}(b)^{\bm{\times}}$};
  \node (e) [below right of=c] {$\mathsf{demote}(a, b, \mathsf{Reader})$};

  \draw[<-] (a) -- (b);
  \draw[<-] (b) -- (c);
  \draw[<-] (c) -- (d);
  \draw[<-] (c) -- (e);

  \coordinate (notes) at ([xshift=1.5cm] a.east);
    \node[note, anchor=west] at (notes |- a) {$a=\mathsf{Admin}$};
    \node[note, anchor=west] at (notes |- b) {$a=\mathsf{Admin}, b=\mathsf{Reader}$};
    \node[note, anchor=west] at (notes |- c) {$a=\mathsf{Admin}, b=\mathsf{Writer}$};
    \node[note, anchor=west] at (notes |- d) {$a=\mathsf{Admin}, b=\mathsf{Reader}$};
    \node[note, anchor=west] at (notes |- e) {$a=\mathsf{Admin}, b=\mathsf{Reader}$};

\end{tikzpicture}
\caption{Admin operations are ordered before Writer operations. As a result, $\mathsf{write}(b)$ is not authorised because $b$ is now a Reader.}
\label{fig:concurrent}
\end{figure}
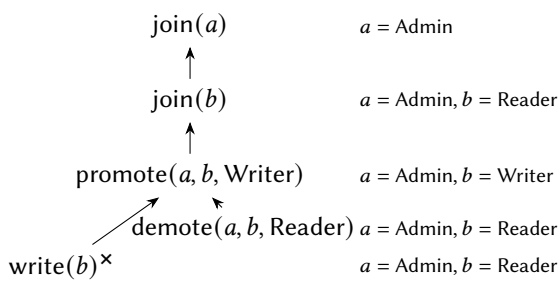

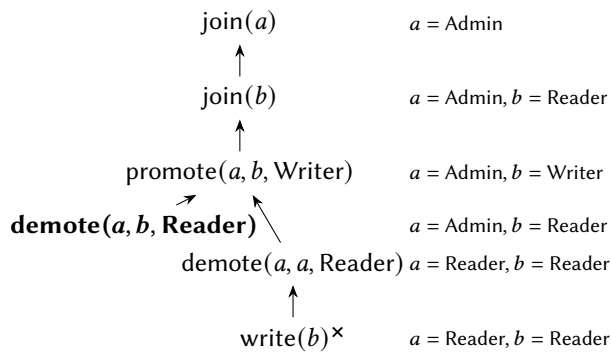
\begin{figure}[h]
\centering
\begin{tikzpicture}[
  node distance=1cm,
  >=Stealth,
  note/.style={
    align=left,
    text width=3cm,
    font=\footnotesize,
  }
]
  \node (a) {$\mathsf{join}(a)$};
  \node (b) [below of=a] {$\mathsf{join}(b)$};
  \node (c) [below of=b] {$\mathsf{promote}(a, b, \mathsf{Writer})$};
  \node (d) [below left of=c, xshift=-7mm] {$\bm{\mathsf{demote}(a, b, \mathsf{Reader})}$};
  \node (e) [below right of=c, yshift=-5mm] {$\mathsf{demote}(a, a, \mathsf{Reader})$};
  \node (f) [below of=e] {$\mathsf{write}(b)^{\bm{\times}}$};

  \draw[<-] (a) -- (b);
  \draw[<-] (b) -- (c);
  \draw[<-] (c) -- (d);
  \draw[<-] (c) -- (e);
  \draw[<-] (e) -- (f);

  \coordinate (notes) at ([xshift=1.5cm] a.east);
    \node[note, anchor=west] at (notes |- a) {$a=\mathsf{Admin}$};
    \node[note, anchor=west] at (notes |- b) {$a=\mathsf{Admin}, b=\mathsf{Reader}$};
    \node[note, anchor=west] at (notes |- c) {$a=\mathsf{Admin}, b=\mathsf{Writer}$};
    \node[note, anchor=west] at (notes |- d) {$a=\mathsf{Admin}, b=\mathsf{Reader}$};
    \node[note, anchor=west] at (notes |- e) {$a=\mathsf{Reader}, b=\mathsf{Reader}$};
    \node[note, anchor=west] at (notes |- f) {$a=\mathsf{Reader}, b=\mathsf{Reader}$};

\end{tikzpicture}
\caption{User $a$ backdated event $\mathsf{demote}(a, b, \mathsf{Reader})$ despite not being authorised to demote anymore. Whether it executes before or after $\mathsf{demote}(a, a, \mathsf{Reader})$ depends solely on attacker-controlled data. This breaks safety property~\ref{prop:safety}.}
\label{fig:selfdemote}
\end{figure}

\emph{Seniority ranking}~\cite{kleppmannKeynoteByzantine}, as proposed by Kleppmann, orders users and is defined as:
\begin{quote}
    Group creator has rank $1$, user added by a user with rank $i$ has rank $i+1$, break ties by lexicographical order on hashes of operations
    that added the users.~\cite{kleppmannKeynoteByzantine}
\end{quote}
Should two concurrent events involve the same user (and thus the same seniority), these events are tie-broken on data within the event,
e.g. timestamp or event hash.
Initially, this appears to enforce the safety property~\ref{prop:safety}, as shown in Figure~\ref{fig:concurrent}.
However, \emph{self-demotions} break the invariant because a malicious user controls all data within the event.
This means self-demoted users can backdate events to re-use permissions they gave up, as shown in Figure~\ref{fig:selfdemote}.

This highlights a tension between the ways concurrent events sent by the same role can be ordered, which are outlined below.

\paragraph{\textbf{User-based arbitration}:}
This means the protocol arbitrarily and consistently decides which user with the same role will win tie-breaks.
Seniority ranking is one such example, where the user who first gained their role will always win.
By sorting all users into a total order, this centralises authority as there is always one user who is guaranteed to win any conflicts, even if the roles indicate shared authority.
If this most senior user is demoted, they can simply backdate a retaliatory demotion to cancel out their own demotion, breaking safety properties ~\ref{prop:safety} and ~\ref{prop:impartial}.
Seniority ordering is incomplete when handling concurrent events sent by the same user, resulting in the inability to enforce self-demotions.

\paragraph{\textbf{Event-based arbitration}:}
This means the protocol chooses based on the revoking event's timestamps or event IDs.
Matrix's state resolution algorithm~\cite{matrixRoomVersion} is one example. This allows users with the same role to both win tie-breaks equally.
Malicious users may game this system to consistently win, e.g. by making superficial modifications to the event data to produce a lower / higher event ID hash.
If two admins behave in this manner, it results in multiple rollbacks whilst they fight each other. This violates safety properties ~\ref{prop:safety} and ~\ref{prop:impartial}.

\paragraph{\textbf{Consensus-based arbitration}:}
Every Admin could propose the order they saw demotions.
The majority would decide the final execution order.
Consensus is vulnerable to Sybil attacks~\cite{douceur2002sybil}, because existing Admins can backdate sockpuppet Admin accounts to gain a majority, breaking safety property ~\ref{prop:impartial}. If user creation could be controlled to ensure one entity cannot have multiple users then this would mitigate this attack.

\paragraph{\textbf{Proof-of-X arbitration}:}
In the seminal Bitcoin paper, Nakamoto summarises that ``Proof-of-work is essentially one-CPU-one-vote''~\cite{nakamoto2009bitcoin}. CPUs are not equally distributed among peers. 
Therefore, Proof-of-Work arbitration approximates user-based arbitration, where the Admin with the more powerful CPU can influence conflict resolution, breaking safety property ~\ref{prop:impartial}.
Proof of Stake systems are inappropriate because CRDTs lack any inherent economic incentive meaning slashing is not a meaningful punishment~\cite{buterin2020incentives}.
Even if such an incentive existed, it would also not be fairly distributed~\cite{celig2025distributional} among Admins,
approximating user-based arbitration, and therefore breaking safety property ~\ref{prop:impartial}.

Backdating exacerbates the difficulty of handling concurrent events because
a demoted Admin can retaliate and turn their sequential demotion into a concurrent conflict. 
Is it possible to detect and perform some remediation upon detecting backdating?
\citeauthor{almeida2024blocklace} propose
that Byzantine peers are excluded from participating after detecting equivocation (backdating)~\cite{almeida2024blocklace}.
However, peers will not agree \textit{which event} was backdated, so some amount of backdating is accepted,
making this solution incomplete.

Consider two scenarios: (1) Admins $a$ and $b$ concurrently demote each other due to genuine network delay, and (2) Admin $b$ receives $a$'s demotion but retaliates by backdating a concurrent demotion.
In scenario (2), the conflict must be resolved in favour of $a$, as $b$'s demotion is retaliatory.
However, the resulting DAGs are structurally identical.
No peer can distinguish these cases from the DAG alone.
Any rule that guarantees $a$ wins in scenario (2) also guarantees $a$ wins in scenario (1), denying $b$ a legitimate outcome.
Seniority ranking sidesteps this by always letting the more senior admin win, but this merely forbids the less senior admin from ever prevailing, even legitimately, while leaving the more senior admin free to backdate.
Because the DAG cannot encode which case is occurring, external information is required.

\section{Concept: Epoch-Resolved Arbitration}

The concept presented in this section acknowledges that a neutral bystander,
a ``finality arbiter'', is required
in order to arbitrate an ordering between duelling admins' events.
This neutral bystander can also prevent
Byzantine peers from ``going back on their word'' by backdating events.
A finality arbiter is a CRDT peer which periodically announces the event IDs of its current \emph{sources} 
in its local DAG.
Other CRDT peers use this information when ordering events into an execution order for calculating the materialised view.
As such, both duelling admins must trust that the finality arbiter is emitting sources honestly.

\subsection{Epoch events}

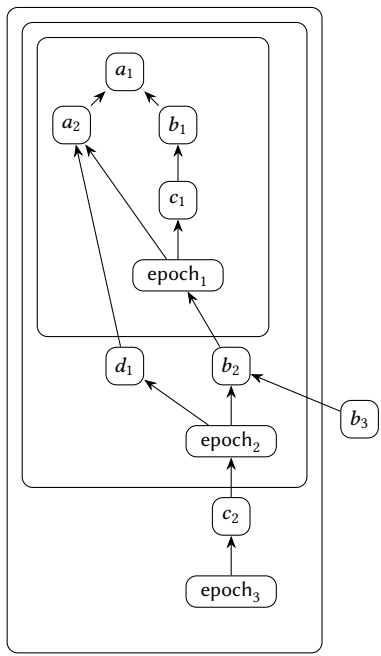
\begin{figure}[t]
\centering
\begin{tikzpicture}[
  >=Stealth,
  node distance=9mm,
  font=\footnotesize,
  msg/.style={
    draw,
    rounded corners,
    minimum size=5mm,
    inner sep=1.5pt
  },
  epochmsg/.style={
    draw,
    rounded corners,
    minimum width=1.2cm,
    inner sep=2pt
  },
  epoch/.style={
    draw,
    rounded corners,
    inner sep=4mm
  },
  outerepoch/.style={
    draw,
    rounded corners,
    inner sep=6mm
  }
]

\node[msg] (a1) {$a_1$};
\node[msg, below right of=a1] (b1) {$b_1$};
\node[msg, below of=b1] (c1) {$c_1$};
\node[msg, below left of=a1] (a2) {$a_2$};

\draw[->] (b1) -- (a1);
\draw[->] (c1) -- (b1);
\draw[->] (a2) -- (a1);

\node[epochmsg, below of=c1] (ae1) {$\mathsf{epoch}_1$};

\draw[->] (ae1) -- (a2);
\draw[->] (ae1) -- (c1);

\begin{pgfonlayer}{background}
  \node[epoch, fit=(a1)(b1)(c1)(a2)(ae1), xshift=2mm, yshift=-2mm] {};
\end{pgfonlayer}

\node[msg, below right of=ae1,yshift=-5mm] (b2) {$b_2$};
\node[msg, below left of=ae1,yshift=-5mm] (d1) {$d_1$};
\node[epochmsg, below of=b2] (ae2) {$\mathsf{epoch}_2$};

\draw[->] (b2) -- (ae1);
\draw[->] (ae2) -- (b2);
\draw[->] (d1) -- (a2);
\draw[->] (ae2) -- (d1);

\begin{pgfonlayer}{background}
  \node[epoch, fit=(b2)(d1)(ae2)(a1)(b1)(c1)(a2)(ae1)] {};
\end{pgfonlayer}

\node[msg, below of=ae2] (c2) {$c_2$};
\node[epochmsg, below of=c2] (ae3) {$\mathsf{epoch}_3$};

\draw[->] (ae3) -- (c2);
\draw[->] (c2) -- (ae2);

\begin{pgfonlayer}{background}
  \node[epoch, fit=(b2)(d1)(ae2)(a1)(b1)(c1)(a2)(ae1)(ae3)(c2), inner sep=6mm,] {};
\end{pgfonlayer}

\node[msg, below right of=b2, xshift=10mm] (b3) {$b_3$};
\draw[->] (b3) -- (b2);
\end{tikzpicture}
\caption{Epochs define a \textit{closed past}~\cite{almeida2024framework} to prevent backdating. Event $b_3$ is in a \textit{pending epoch}, despite it pointing to an earlier event in a previous epoch.}
\label{fig:epochonion}
\end{figure}

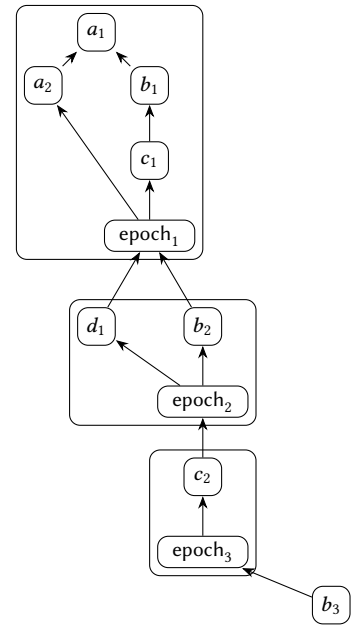
\begin{figure}[t]
\centering
\begin{tikzpicture}[
  >=Stealth,
  node distance=9mm,
  font=\footnotesize,
  msg/.style={
    draw,
    rounded corners,
    minimum size=5mm,
    inner sep=1.5pt
  },
  epochmsg/.style={
    draw,
    rounded corners,
    minimum width=1.2cm,
    inner sep=2pt
  },
  epoch/.style={
    draw,
    rounded corners,
    inner sep=4mm
  },
  outerepoch/.style={
    draw,
    rounded corners,
    inner sep=6mm
  }
]

\node[msg] (a1) {$a_1$};
\node[msg, below right of=a1] (b1) {$b_1$};
\node[msg, below of=b1] (c1) {$c_1$};
\node[msg, below left of=a1] (a2) {$a_2$};

\draw[->] (b1) -- (a1);
\draw[->] (c1) -- (b1);
\draw[->] (a2) -- (a1);

\node[epochmsg, below of=c1] (ae1) {$\mathsf{epoch}_1$};

\draw[->] (ae1) -- (a2);
\draw[->] (ae1) -- (c1);

\begin{pgfonlayer}{background}
  \node[epoch, fit=(a1)(b1)(c1)(a2)(ae1), inner sep=1mm] {};
\end{pgfonlayer}

\node[msg, below right of=ae1,yshift=-5mm] (b2) {$b_2$};
\node[msg, below left of=ae1,yshift=-5mm] (d1) {$d_1$};
\node[epochmsg, below of=b2] (ae2) {$\mathsf{epoch}_2$};

\draw[->] (b2) -- (ae1);
\draw[->] (ae2) -- (b2);
\draw[->] (d1) -- (ae1);
\draw[->] (ae2) -- (d1);

\begin{pgfonlayer}{background}
  \node[epoch, fit=(d1)(ae2)(b2), inner sep=1mm] {};
\end{pgfonlayer}

\node[msg, below of=ae2] (c2) {$c_2$};
\node[epochmsg, below of=c2] (ae3) {$\mathsf{epoch}_3$};

\draw[->] (ae3) -- (c2);
\draw[->] (c2) -- (ae2);

\begin{pgfonlayer}{background}
  \node[epoch, fit=(ae3)(c2), inner sep=1mm] {};
\end{pgfonlayer}

\node[msg, below right of=ae3, xshift=10mm] (b3) {$b_3$};
\draw[->] (b3) -- (ae3);
\end{tikzpicture}
\caption{The partial execution order for events in Figure \ref{fig:epochonion} after ordering by epoch. After sorting by epoch, events can then be sorted by revocation, timestamp, event hash, etc.}
\label{fig:epochexecution}
\end{figure}

A finality arbiter announces its current sources via \textit{epoch events}.
Assuming the finality arbiter does not backdate epoch events, this forms onion-like layers of events in each epoch,
as shown in Figure~\ref{fig:epochonion}.
Events are then ordered by their epoch \textit{first}, thus finalising the \textit{execution order}
of the CRDT, as shown in Figure~\ref{fig:epochexecution}.
For example, $d_1$ is not part of the first epoch, despite it pointing to an event $a_2$
within the first epoch. In the execution order, this means $\mathit{epoch}_1 \rightarrow d_1$.
Events not in any epoch are in a \textit{pending epoch} which executes after all other epochs,
regardless of what the causal predecessors are in the event itself.
Epochs are triggered based on the number of events in the pending epoch, but could also be triggered on-demand
for non-monotonic events such as demotions.

\section{Discussion}

The motivation behind epochs is to achieve finality for CRDT state in order to meet safety property~\ref{prop:impartial}.
The Duelling Admins problem is a manifestation of the lack of finality in CRDT state.
This means epochs only need to be sent in response to non-monotonic operations such as the $\mathsf{demote}$ operation,
which have the possibility to roll back events.
ERA resolves the Duelling Admins problem by providing a central point of coordination via a finality arbiter. 
Finality arbiters also have applications beyond group management CRDTs.
Any situation where the materialised view is used to perform non-monotonic operations at the application level may benefit from finality.
Traditional consensus requires coordination \emph{before} operations commit~\cite{brewer2012cap}.
In contrast, ERA enables operations to be applied \emph{before} coordination, with finalised operations being communicated via epoch events.
This is similar to how finality works in blockchains, as miners do not coordinate to mine blocks but instead gradually converge on a single fork.
The trade-off is that, until finality is reached, both ERA and blockchains can exhibit rollbacks.

A delegation chain could authorise finality arbiters to sign each event to communicate finality, instead of sending epoch events.
However, epoch events allow large batches of events to be finalised succinctly, and ensures that the finalised events respect causal ordering. Individually signing events scales poorly in comparison, and opens up new failure modes where an event could be finalised before its causal predecessors.

The consistency of finalised events relies on the
monotonic creation of epochs by finality arbiters.
This results in two consequences outlined below.

\subsection{Arbiter Trust}

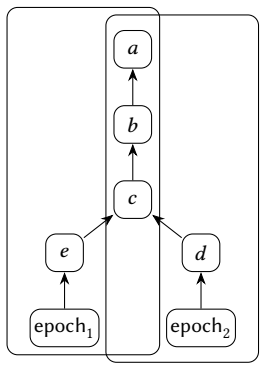
\begin{figure}[h]
\centering
\begin{tikzpicture}[
  >=Stealth,
  node distance=1cm,
  font=\footnotesize,
  msg/.style={
    draw,
    rounded corners,
    minimum size=5mm,
    inner sep=1.5pt
  },
  epochmsg/.style={
    draw,
    rounded corners,
    minimum width=1.2cm,
    inner sep=2pt
  },
  epoch/.style={
    draw,
    rounded corners,
    inner sep=4mm
  },
  outerepoch/.style={
    draw,
    rounded corners,
    inner sep=6mm
  }
]

\node[msg] (a) {$a$};
\node[msg, below of=a] (b) {$b$};
\node[msg, below of=b] (c) {$c$};

\node[msg, below left of=c, xshift=-2mm] (e) {$e$};
\node[msg, below right of=c, xshift=2mm] (d) {$d$};
\node[msg, below of=d] (epoch2) {$\mathsf{epoch}_2$};
\node[msg, below of=e] (epoch1) {$\mathsf{epoch}_1$};

\draw[->] (b) -- (a);
\draw[->] (c) -- (b);
\draw[->] (d) -- (c);
\draw[->] (e) -- (c);
\draw[->] (epoch1) -- (e);
\draw[->] (epoch2) -- (d);

\begin{pgfonlayer}{background}
  \node[epoch, fit=(a)(b)(c)(e)(epoch1), inner sep=2mm, xshift=-1mm, yshift=1mm] {};
\end{pgfonlayer}

\begin{pgfonlayer}{background}
  \node[epoch, fit=(a)(b)(c)(d)(epoch2), inner sep=2mm, xshift=1mm] {};
\end{pgfonlayer}
\end{tikzpicture}
\caption{Concurrent epochs can cause history to deterministically re-order events $d$ and $e$.
If $\mathsf{epoch}_1 < \mathsf{epoch}_2$ then $\{a, b, c, e\} \rightarrow d$.
If $\mathsf{epoch}_2 < \mathsf{epoch}_1$ then $\{a, b, c, d\} \rightarrow e$.}
\label{fig:epochconcurrent}
\end{figure}

The finality arbiter is trusted to monotonically generate epochs.
The arbiter can ``rewrite history'' if it sends concurrent epoch events, as shown in Figure~\ref{fig:epochconcurrent}: a form of \emph{detectable backdating}.
To detect this, epoch events must be signed by the arbiter.
Whilst the arbiter can misbehave by signing concurrent epoch events,
their signatures allow other peers to prove misbehaviour and enforce accountability, 
similar to certificate authorities in a public key infrastructure.
Techniques like transparency logs~\cite{transparencyLogsCertificate} can also be applied to finality arbiters as well.
For example, a peer holding both concurrent epoch events and the relevant DAG history could construct a fraud proof demonstrating misbehaviour.
The cost of verifying such a proof depends on how much of the DAG must be replayed, which remains an open question.
Nonetheless, signed epoch events provide a foundation for accountability.

The protocol may also define an ordered list of finality arbiters, where each arbiter independently issues epoch events but the protocol only uses the first arbiter.
Upon being provided a fraud proof, the second arbiter is used, and so on.
This does not prevent history from being rewritten as there is no guarantee that each arbiter will agree on which events are in which epoch,
but it does prevent a malicious finality arbiter from manipulating CRDT state to its advantage.

The finality arbiter could be a \emph{mutually trusted} third party unaffiliated with any group participants. Group participants can then encrypt the underlying CRDT state to keep it confidential from the arbiter.
Encryption would not prevent history from being rewritten, but it would make it much more difficult for a malicious finality arbiter to intentionally rewrite history in a way which is beneficial, as it would need to distinguish between ciphertexts.
Protocols like Keyhive already encrypt CRDT events and so can more freely delegate responsibility for ordering events to third parties.

For CRDTs which operate on the public internet, lists of publicly available finality arbiters could be published along with uptime percentages and the number of times that the arbiter has been caught backdating.
This would provide a mechanism for users to find and select reliable third party finality arbiters.
Trust in finality arbiters does \emph{not} extend to other CRDT group participants.
With a trusted finality arbiter, all properties hold even if the creator is Byzantine
because the group creator cannot override the ordering provided by the finality arbiter.
This requires participants to trust the finality arbiter that the group creator has selected.

\subsection{Epoch Reliability} \label{reliability}

The finality arbiter must reliably receive events, generate epoch events, and transmit them to other peers.
If this does not happen, finality will not be reached on a growing set of events.
Despite this, peers are still able to communicate with each other, preserving liveness property~\ref{prop:liveness}.
The absence of finality would enable backdating attacks outlined in this article on events in the \emph{pending epoch} only.
The finality arbiter should have high uptime and be positioned on the network to maximise reachability to ensure it has the best chance to receive events and transmit epoch events to other peers.

\subsection{Group Creator as Finality Arbiter}

The group creator is the first Admin in the group.
By virtue of being the first, they are able to backdate to an earlier point in time than anyone else.
Seniority ranking formalises this idea by ordering all users by delegation chain length.
Similarly, it is possible to formalise this idea in ERA by using the group creator as the finality arbiter.
This grants the group creator the ability to win any duels, placing them in a unique ``Creator'' role which is more powerful than Admin.
P2P networks would struggle to use the group creator as the finality arbiter because end-user devices have poor reliability, as discussed in Section~\ref{reliability}.
However, federated systems where peers are servers which are identified via domain may fare better.
Matrix recently~\cite{matrixRoomVersion} introduced a new ``Creator'' role.
This role is assigned to one or more users in the first event in the DAG.
Critically, Creator role assignments are immutable. This means it is impossible for creators to demote each other or themselves. This makes ``Duelling Creators'' impossible.
This opens up the possibility for the creator to fulfil the role of finality arbiter without introducing trust issues because the creator is already trusted.
The primary disadvantage of this approach is that the creator's server may not be reliable.
In Matrix, most Admins reside on only 1 domain.
Figure~\ref{fig:domains} shows the privileged user distribution of \emph{unique} domains for a large representative sample of public and private rooms (groups) on the matrix.org homeserver, one of the largest Matrix servers federating with 40,779 other servers.
Aggregate data was collected on 11 March 2026 on a random subset of rooms accessible to the matrix.org server (N=9048520).
After excluding rooms with less than 3 privileged users, defined as having a power level > 0, N=213490 rooms were included.
The distribution approximates a geometric decay, consistent with the connectivity patterns commonly observed in networked systems~\cite{scaling}.
Consequently, quorum-based solutions would not materially improve reliability in Matrix's federated ecosystem because that needs at least 3 privileged domains.
Instead, should the group creator's server go permanently offline, the room should be upgraded to transfer ownership to a new user.

\begin{figure}[htbp]
    \centering
    \begin{tikzpicture}
    \begin{axis}[
        ybar,
        xlabel={Unique privileged domains per room},
        ylabel={Percentage of rooms},
        symbolic x coords={1, 2, 3, 4, 5, 6, 7+},
        xtick=data,
        nodes near coords={\pgfmathprintnumber\pgfplotspointmeta\%},
        bar width=20pt,
        fill=black!20,
        draw=black,
        ymin=0, ymax=100,
    ]
    \addplot[fill=black!20, draw=black] coordinates {
        (1, 62.4)
        (2, 23.0)
        (3, 8.9)
        (4, 2.6)
        (5, 1.2)
        (6, 0.6)
        (7+,1.3)
    };
    \end{axis}
    \end{tikzpicture}
    \caption{Domain distribution of privileged users e.g. Admins}
    \label{fig:domains}
\end{figure}

\section{Conclusion and Future Work}

We discussed the Duelling Admins problem: if two admins concurrently revoke each other, what should be the outcome?
Different solutions were examined and we proposed our own ERA solution using a finality arbiter.
\emph{Finality arbiters} provide a technique to solve the Duelling Admins problem in a fair way (property ~\ref{prop:impartial}) which guarantees revocation cycles are broken.
They guard against undesirable rollbacks and backdating attacks by finalising the execution order of events.
If the finality arbiter is unreachable, finalised events are safe from being rolled back, but no other events will be finalised.
This provides application designers a choice: security sensitive operations may wait for finality (and thus may block) for some operations to enforce safety properties ~\ref{prop:safety} and ~\ref{prop:impartial},
whilst other operations may occur immediately with the risk of invalidation to provide liveness property ~\ref{prop:liveness}. This is a form of mixed consistency~\cite{shapiro2018justrightconsistencyreconcilingavailability}.

A drawback to this approach is that the finality arbiter is trusted to order events in a timely, unequivocal manner.
Finality arbiters may intentionally delay ordering events or may intentionally re-order events, the latter being a form of detectable backdating.
Two main techniques can be used to disincentivise malicious behaviour.
The first is to encrypt CRDT event data from the finality arbiter such that a non-colluding finality arbiter cannot intentionally re-order events.
The second is to attach a public reputation to the finality arbiter such that malicious behaviour is socially penalised.
Alternatively, we discussed an additional ``Creator'' role which can act as a finality arbiter to circumvent the need to have an external arbiter.

This design separates finalised and unfinalised events.
Future research may take advantage of the separation by having novel algorithms for finalised events vs. temporarily unfinalised events. Finalised events could possibly be compacted or garbage collected.
It may also be possible to use finality to gradually mutate the list of finality arbiters over time to provide a mechanism
to remove Byzantine finality arbiters.

\begin{acks}
I thank Martin Kleppmann for helpful discussions, Florian Jacob for detailed feedback on earlier drafts, and the anonymous PaPoC reviewers who brought this paper to completion.
The author is employed by Element Creations Ltd, which develops software based on the Matrix protocol.
\end{acks}

\balance{}
\bibliographystyle{ACM-Reference-Format}
\bibliography{references}

\end{document}